\documentclass{article}
            \usepackage{graphicx}
            \usepackage{amsmath,amssymb}
\begin{document}
            \title{The no information at a distance principle and local mathematics: some effects on physics and geometry}
          \author{Paul Benioff\\
            Physics Division, Argonne National
           Laboratory,\\
           Argonne, IL 60439, USA \\
           email: pbenioff@anl.gov}
           \maketitle
           
            \begin{abstract}
            Local mathematics assumes the existence of  number structures of different types,  vector spaces, etc. localized at each space time point. Relations between number structures  at different locations are based on two aspects: distinction between two so far conflated concepts, number and number value and the "No information at a distance" principle. This principle forbids the choice of the value of a number at one location to determine the value of the same number at another location. Value changing connections, related to a real valued  field, $g,$ move numbers between structures at different locations.  The effect of the $g$ field, or its exponential equivalent,  $g(y)=e^{\alpha(y)},$ on numbers extends to other mathematical structures,  vector spaces, etc.
           The presence of $\alpha$ affects theoretical descriptions of quantities in physics and geometry.  Two examples are described, the effect on the Dirac Lagrangian in gauge theory, and the effect on path lengths and distances in geometry. The gradient field of $\alpha$, $\vec{A},$ appears in  the Lagrangian as a spin $0$, real scalar field that couples to the fermion field.  Any value for the mass of $\vec{A}$ is possible. The lack of direct experimental evidence for the presence of the $g$ or $\alpha$ field means that the  field must be essentially constant within a local region of the cosmological universe.  Outside the local region there are no restrictions on the field. Possible physical candidates, (inflaton, dark matter, dark energy) for $\alpha$ are noted.
           \end{abstract}

           \section{Introduction}

           The relation between the foundations of mathematics and physics is a subject of much interest. This is well expressed by Wigner \cite{Wigner} and others \cite{Omnes,Plotnitsky,Hitchin,Dirac,Von Neumann,WikiRMP} in papers on the unreasonable, and reasonable effectiveness of mathematics in  science.  This effectiveness appears to be  a problem, especially if one accepts the Platonic views of mathematical elements as  having an ideal existence outside of space and time. From this perspective it is not understandable why mathematics should have anything to do with the description of physical systems that move and interact inside a space time arena.  This leads to the wish for a coherent theory of physics and mathematics together as one coherent whole rather than as two separate disciplines \cite{BenCTPM1,BenCTPM2}.

           One approach to this problem is to replace global mathematics existing outside of space and time with local mathematics.  In this approach mathematical systems of different types are considered to be  structures \cite{Shapiro}. Here these  structures are considered to be local with   separate structures of each type  existing at each point of space time.  Relations or maps between systems of different types correspond to maps or relations between structures of different types at the same or different locations in space time.   This puts mathematical structures and physical systems on a more  equal footing in that they both exist in space and time.

           The idea of local mathematical structures has a precedent in that gauge theories are based on vector spaces localized at different points of space time \cite{Montvay}. To each point, $x,$ is associated a vector space, $\bar{V}_{x}.$

           Relations between states in vector spaces at different locations are influenced by the "No information at a distance" principle \cite{Montvay,Mack}. This principle says that information used to choose a basis set of states in $\bar{V}_{x}$ cannot influence the choice of a basis set of states  in another vector space, $\bar{V}_{y},$ at a different location, $y$. This condition leads to the existence of unitary gauge transformations between states at different locations. These transformations, first investigated by Yang Mills \cite{YangM}  for nonAbelian gauge theories,  led to  the  description of the standard model for elementary particles in physics.

           An aspect of this that seems strange is that, as mathematical structures, vector spaces do not exist by themselves.  They are closely associated with scalar fields such as the real or complex numbers.  The association of local vector spaces with a single global scalar field outside of space time seems problematic \cite{BenNOVA}.

           For this reason it seemed worthwhile  to replace a single global scalar field  by separate local scalar fields. These fields,  as structures for numbers of different types, become the scalars for the vector spaces, both at the same location \cite{Ben1st}.

           This leads directly to the idea that if number structures of different types are local, this locality should extend to  mathematical structures for all types of systems that include numbers in their axiomatic descriptions.  There are many of these types of structures in mathematics. Group representations, algebras, matrices, all types of vector spaces, are examples of this type of structure.

           An aspect of number structures that is relevant here is the observation that the concept of number is different from that of number value. For each type of number, there exist many structures in which the same number has different values. For real numbers there is an infinity of real number structures, one for each real number. For complex number structures there is an infinity of structures, one for each complex number.

           These structures are referred to as scaled structures with associated real or complex scaling or value factors.  These scaled structures have  one base set of numbers in common.  With one exception, the numbers in the base set have no intrinsic value.  The value of a number is determined by the scaling factor and the position of the number in the structure being considered. The exception is the number $0.$ Its value is $0$ in number structures for all scaling factors.

           The concept of scaling  for number structures is easily extended to structures of all  types that include numbers in their description.  For example, one can have scaled vector spaces, scaled algebras, etc. Most of the work done so far has been limited  to numbers of different types and to vector spaces.

           This distinction between number and number value and the resulting existence of number structures with different scaling factors,  is an example of the distinction between Diophantine and nondiophantine arithmetic \cite{BurginDio,BurginDio2}. The number structure with a scaling factor of $1$ would correspond to the usual or Diophantine arithmetic.  The structures with scaling factors different from $1$ correspond to nondiophantine arithmetics.  The description of  natural number structures with different scaling factors  given in this work is an example of projective arithmetics \cite{Burginpro}.

            As defined here the scaling factors are linear in that they preserve order relations between numbers.  A more general functional type of scaling has been used by Czachor \cite{Czachor1} to suggest that the relativity of arithmetic may be a symmetry of physics. It was also suggested that this relativity may be a source of dark energy \cite{Czachor2}.  Details of the use of this more general type of scaling for  physics and geometry remain to be investigated.

            The description used in gauge theories to describe  relations  between vector spaces at different locations by unitary gauge transformations is extended here to number structures. The "No information at a distance" principle \cite{Mack,Montvay} forbids the use of the informational choice of the value of a number at one location to determine the value of the same number at a different location. In other words, the choice of the value of a number at one location does not determine the value of the same number at a different location.

           This leads to a location dependent valuation of number structures. The valuation is implemented  by the introduction of  a scalar scaling or value field, $g.$ This field associates a scaling value $g(y)$ to the local  number structures at $y$. The relation between numbers and number structures at different locations is determined by a parallel transform or connection that maps numbers and structures at one location to those of another. The connection depends on the values of $g$ at the different locations. The connection is the number equivalent of gauge transformations in gauge theory.

           As might be expected, the presence of the $g$ field affects theoretical descriptions of many properties of systems in physics and geometry. This is especially the case for properties described by integrals or derivatives over space or space time.  Reconciliation of these theory predictions with experiment results places restrictions on the $g$ field.

           This work has been described in detail elsewhere \cite{BenNOVA}, \cite{BenINTECH}-\cite{Ben2nd}.  Here two examples, one from physics and the other from geometry, will be given to illustrate the effects of local mathematics and the presence of the $g$ field. The next section is a brief outline of the  effect of scaling on numbers and on other mathematical structures.    The following two  sections  describe the effects of $g$ field on gauge theories and geometry.  It will be seen that presence of the $g$ field introduces a new scalar field  into the Dirac Lagrangian. The field also affects distances between points.  The final section describes restrictions on the $g$ field imposed by experiment.

           \section{Number scaling}\label{NS}
           \subsection{Natural numbers}
           The description of number scaling begins with the idea that the concept of number is separate from that of number value.  The best way to see this is to consider the natural numbers, \begin{equation}\label{0-4etc} 0\;\;1\;\;2\;\;3\;\;4\cdots=N_{1}. \end{equation} Here, as symbols, $0,1,2,3,4$ are names for numbers.\footnote{Each of the numbers along with its name  is a named set \cite{BurginNamed}}.   From now on the distinction between name and number will be dropped.

           The numbers  in Eq. \ref{0-4etc} are well ordered with  $0$ the first, $1$ the second, $2$ the third, and so on in the well ordering.  The value of a number is determined by its position in the well ordering. Thus $0$ has value $0$, $1$ has value $1$, $2$ has value $3$ and so on.

           Consider the subset $$N_{2} =0,2,4,6\cdots$$ of even numbers.  This set inherits the well ordering of the numbers in Eq. \ref{0-4etc}. $0$ is first $2$ is second, $4$ is third, $6$ is fourth and so on. The values of the numbers in this set are determined by their position in the well ordering.  Thus $0$ has value $0$, $2$ has value $1,$ $4$ has value $2$, $6$ has value $3,$ and so on.

           This shows that, with the exception of the number $0$, the numbers have no intrinsic value.  Their values are determined by the set containing them.  The number $2$ has value two in the set of Eq. \ref{0-4etc}. It has value $1$ in the subset, $0,2,4,6$. The number $4$ has value $4$ in Eq. \ref{0-4etc}.  It has value $2$ in the subset.

           This description can be extended to subsets, $N_{n}$ that contain every $nth$ number in $N_{1}$.  For $n =4$ the number $4$ can have three possible values, $4$ in $N_{1}$, $2$ in $N_{2}$, and $1$ in $N_{4}.$ The number of number values a number can have is determined by the prime factors of the number.  If the number is a prime number it has just two values. If it is not a prime number it can have more values.  For example the possible values of the number $30$ can be expressed by the equations \begin{equation}\label{30}30_{1}=15_{2}=10_{3}=6_{5}=5_{6}=3_{10}=2_{15}=1_{30}.\end{equation}  Expressions such as $10_{3}$ mean that the value of the number is $10$ in $N_{3}.$ This representation of numbers and their values will be much used from now on.

           This simple example of the effects of separation of number from number value extends to the other number types, integers, rational, real, and complex numbers. For rational, real, and complex numbers, The base sets, $B_{Ra}$, $B_{R}$ and $B_{C}$ remain unchanged under scaling.  Subsets are not needed.  This is a result of the observation that, unlike the natural numbers, these number types are closed under division.
           \subsection{Real and complex numbers}
           In what follows it is useful to consider mathematical systems of different types as structures \cite{Shapiro} satisfying a set of relevant axioms. Relations between structures are included in the axiomatic descriptions. The closure of vector spaces under scalar multiplication is an example. Here the description is limited to real and complex  numbers as these are most relevant to physics and geometry.  The structures for real and complex numbers are given by \begin{equation}\label{NRNC}\bar{R}=\{B_{R},\pm,\times,\div,<,0,1\}\mbox{ and } \bar{C}=\{B_{C},\pm,\times,\div,^{*},0,1\}.\end{equation}In these definitions $B_{R}$ and $B_{C}$ are base sets, $\pm,\times,\div$ are the basic field operations,$<$ is an order relations, $^{*}$ denotes complex conjugation, and $0$ and $1$ are constants. The structures satisfy axioms for real and complex numbers.

           Let $r$ and $c$ denote real and complex  number values. Real and complex number structures scaled by $r$ and $c$ are given by \begin{equation}\label{NRNCrc}\bar{R}_{r} =\{B_{R},\pm_{r},\times_{r},\div_{r},<_{r},0_{r},1_{r}\}\mbox{ and }\bar{C}_{c}=\{B_{C},\pm_{c},\times_{c},\div_{c},(^{*})_{c},0_{c},1_{c}\}.\end{equation}In these structures $1_{r}$ and $1_{c}$ are numbers in $\bar{R}^{r}$ and $\bar{C}^{c}$ with value $1.$ Also $0_{r}=0=0_{c}$. Let $Op$ denote any one of the four binary field operations, $+,-,\times,\div.$ Then $Op$ denotes an operation value and $Op_{r}$ and $Op_{c}$ denote the operations in $\bar{R}^{r}$ and $\bar{C}^{c}$ with value, $Op.$  Note that for multiplication and division, if $r\neq s$, then $Op_{r}$ differs from $Op_{s}$ by a scaling factor.\footnote{Since rational numbers, integers and the natural numbers are substructures of the real numbers, structures for these number types scaled by real numbers can be defined. For example the set of natural numbers scaled by $\pi$ is given by $$\bar{N}_{\pi}=\{B_{N},+_{\pi},\frac{1}{\pi}\times_{\pi},<_{\pi},0_{\pi},1_{\pi}\}.$$}

\begin{figure}[h!]\vspace{-2cm} \hspace{1cm}{\includegraphics[scale=.5]{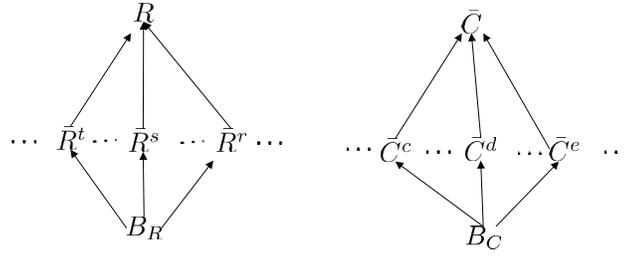}}\vspace{-2cm}\caption{Representations of real and complex number structures for all nonzero respective real and complex scaling factors. Three out of an infinite number of structures are shown for both real and complex numbers.  The base sets, $B_{R}$ and $B_{C}$, and the number value structures, $\bar{R}$ and $\bar{C}$, are the same for all scaled real and complex number  structures .} \label{BS2.1}\end{figure}

           The fact that $r$ and $c$ can be any real and complex number values  leads to collections of scaled structures, one for each scaling factor.  The base sets remain the same for all scaling factors.  These collections and the real and complex number value structures for real and complex numbers  are illustrated in Figure \ref{BS2.1} as  collections over the two common base sets.  Real and complex number value structures, $\bar{R}$ and $\bar{C},$ which are the same for all the scaled structures, are included in the figure.

           The collections can  be defined by\begin{equation}\label{FRFC} F_{R}=\bigcup_{t}\bar{R}^{t}\mbox{ and }F_{C}=\bigcup_{c}\bar{C}^{c}\end{equation} for all real and complex scaling factors, $t$ and $c$, different from $0$. The base sets are included in the definitions.

            There are two types of maps on $F_{R}$ and $F_{C}$.  One  type is number changing and value preserving.  The other is number preserving and value changing. For each real number, $s$ let $W_{s}$ be a number changing value preserving map on $F_{R}.$ The action of $W_{s}$  is defined by
            \begin{equation}\label{WsR}W_{s}\bar{R}^{t}=\bar{R}^{ts}.\end{equation}If $s$ is negative, the direction of the order relation in $\bar{R}^{ts}$ is changed from that in $\bar{R}^{t}.$ The map is number changing value preserving because $1_{t}$ is a different number from $1_{ts}.$ Both numbers have the same value in their respective structures.  Note that the maps $W_{s}$ for all $s\neq 0$ form a commutative group.  The definition of $W_{c}$  on $F_{C}$ for $c\neq 0$ is defined in a similar fashion.

            The definition of the number preserving value changing map is more complex.  Let $s$ be a nonzero real number.  Define $Z_{s}$ to be the associated number preserving value changing map. One has \begin{equation}\label{ZsRts}Z_{s}\bar{R}^{t}=\bar{R}^{t}_{u}\end{equation} where $u=ts.$ The components of $\bar{R}^{t}_{u}$ are given by \begin{equation}\label{Ztts}\begin{array}{c}Z_{s}(a_{t})=(\frac{t}{u}a)_{u},\;\;Z_{s}(+_{t})=+_{u},\;\; Z_{s}(\times_{t})=(\frac{u}{t}\times)_{u},\\\\Z_{s}(\div_{t})=(\frac{t}{u}\div)_{u},\;\;Z_{s}(<_{t})=<_{u}\mbox{ for }s>0\\\\ Z_{s}(0_{t})=0_{u},\;\;Z_{s}(1_{t}) =(\frac{t}{u}1)_{u}.\end{array}\end{equation} These definitions can be combined to write $\bar{R}^{t}_{u}$ as a structure,
             \begin{equation}\label{Rtts}\bar{R}^{t}_{u}=\{B_{R},\pm_{u}, (\frac{u}{t}\times)_{u},(\frac{t}{u}\div)_{u},<_{u},0_{u},(\frac{t}{u}1)_{u}\}.\end{equation} $\bar{R}^{t}_{u}$ is a representation of the components of $\bar{R}^{t}$ in terms of those in $\bar{R}^{u}$.  Scaling of the division and multiplication operations  is necessary to preserve axiom validity under the action of $Z_{s}.$\footnote{ As an example of axiom preservation,  $((t/u)1)_{u}$ satisfies the multiplicative identity axiom. From $$(\frac{t}{u}a)_{u}(\frac{u}{t}\times)_{u}(\frac{t}{u}1)_{u}=(\frac{t}{u}a)_{u}$$ one obtains $$a_{u}\times_{u}1_{u}=a_{u}$$ by multiplying out the ratios of  scaling factors.} Note that $((t/u)1)_{u}$ is the same base set number in $\bar{R}^{u}$ as is $1_{t}$ in $\bar{R}^{t}.$  Its value differs from $1$ by the factor, $t/u.$ As was the case for $W_{s},$ the maps, $Z_{s},$ for all nonzero $s$ form  a commutative group.

            For complex numbers the action of $Z_{c}$ for $c$ complex is given by \begin{equation}\label{ZcCd}Z_{c}\bar{C}^{d}=\bar{C}^{d}_{e}\end{equation}where $e=cd.$ The components of $\bar{C}^{d}_{e}$ are given by Eq. \ref{Ztts} with the replacement of $Z_{s}(<_{t})$ with $$Z_{c}((a_{d})^{*_{d}}) =(\frac{d}{e})_{e}(a_{e})^{*_{e}}=(\frac{d}{e}(a^{*}))_{e}.$$  These components can be combined to define a structure $\bar{C}^{d}_{e}$ by \begin{equation}\label{Cdcd}\bar{C}^{d}_{e}=\{B_{C},\pm_{e},(\frac{e}{d}\times)_{e}, (\frac{d}{e}\div)_{e}, (\frac{d}{e})_{e}(-)^{*_{e}},0_{e},(\frac{d}{e}1)_{e}\}.\end{equation} The blank in the complex conjugation component denotes a number in $\bar{C}^{e}.$

            It must seem strange that, with both $d$ and $e$ complex,  $((d/e)1)_{e}$ is both the identity and is a real number.\footnote{Details are given in \cite{Ben2nd}}  This is an illustration of the fact that properties of numbers do not exist in the abstract, independent of structure membership.  Their properties are determined by the structure containing them.

             It should also be emphasized that  operations such as multiplication and division  are carried out on numbers before their  transport from $\bar{C}^{d}$ to $\bar{C}^{d}_{e},$ not after. Implementing  the operations in the target structures on transported numbers gives a different result.  For example, for any complex number value, $a$, $$(\frac{d}{e}a)_{e}\times_{e}(\frac{d}{e}1)_{e}\neq (\frac{d}{e}a)_{e}$$  and $$Z_{c}((a_{d})^{*_{d}})= (\frac{d}{e}(a)^{*})_{e}\neq ((\frac{d}{e}a)_{e})^{*_{e}}.$$

             \subsection{Vector spaces}
             As noted before, number scaling extends to all mathematical structures whose description includes numbers.  Vector spaces are an example.  They can be based on either real or complex numbers as the scalar field. Here the scalar field will be assumed to be complex.  Spaces based on real scalar fields are a simplification in  that the real scalars are a subfield of the complex scalars. However the scaling factor will be restricted to be a real number value.

             For each nonzero real  number value $r$ let $\bar{V}^{r}$ be a scaled normed vector space.  It is defined by
             \begin{equation}\label{bVr}\bar{V}^{r}=\{B_{V},\pm_{r},\cdot_{r}, |-|_{r}, \psi_{r}\}.\end{equation} Here $\psi_{r}$ denotes an arbitrary vector with vector value, $\psi$, $|-|_{r}$ denotes the norm of a vector value, $\cdot_{r}$ is scalar vector multiplication, and $\pm_{r}$ is  linear superposition. The values of these three operations are represented by $|-|$, $\cdot,$ and $\pm.$  The base set of vectors is represented by $B_{V}.$ The associated scalar field is $\bar{C}^{r}.$ Note that one can express the norm in the form $|\psi_{r}|=|\psi|_{r}.$ This is a real number in $\bar{C}^{r}.$

             The collection of vector spaces and associated complex number fields for all real nonzero scaling factors is represented by \begin{equation}\label{FVC}
              F_{VC}=\bigcup_{r}(\bar{V}^{r}\times\bar{C}^{r}). \end{equation} There is one base set pair, $B_{V}\times B_{C}$ and one pair of value structures, $\bar{V}\times\bar{C}$ for the collection of scaled spaces and fields. As was the case for numbers there are two types of maps on $F_{VC}$. One is vector and number changing, vector and number value preserving, and  the other is vector and number preserving and vector value and number value changing.   The first type is denoted by $W_{p}$ where \begin{equation}\label{WpVrCr}W_{p}(\bar{V}^{r}\times\bar{C}^{r})=\bar{V}^{rp}\times\bar{C}^{rp}.\end{equation}  As was the case for scalars The collection of $W_{p}$ for all nonzero $p$ form a commutative group.

              The description of the map $Z_{p}$ for number structures can be extended to vector spaces.  Define the actions of $Z_{p}$ on the components of $\bar{V}^{r}$ by \begin{equation}\label{Zppsi}\begin{array}{c}Z_{p}(\psi_{r})=(\mbox{\Large$\frac{r}{q})_{q}$}\psi_{q}=(\mbox{\Large$\frac{r}{q}$}\psi)_{q},\;\; Z_{p}(\pm_{r})=\pm_{q} \\\\Z_{p}(\cdot_{r})=(\mbox{\Large$\frac{q}{r}$}\cdot)_{q},\;\; Z_{p}(|\psi|_{r})=(\mbox{\Large$\frac{r}{q}$}|\psi|)_{q},\;\; Z_{p}(\psi_{r})=(\mbox{\Large$\frac{r}{q}$}\psi)_{q}.\end{array}\end{equation} Here $q=pr.$

              This definition can be combined with that for $\bar{C}^{r}_{q}$ to write \begin{equation}\label{ZpCrVr}Z_{p}(\bar{C}^{r}\times\bar{V}^{r}) =\bar{C}^{r}_{q}\times\bar{V}^{r}_{q}\end{equation} where \begin{equation}\label{bVrq}\bar{V}^{r}_{q}=\{B_{V},\pm_{q},(\frac{q}{r}\cdot)_{q}, (\frac{r}{q}|\psi|)_{q},(\frac{r}{q}\psi)_{q}\}.\end{equation}

              Note that the norm is special in that \begin{equation}\label{frqpsi}(\frac{r}{q}|\psi|)_{q}=|\frac{r}{q}\psi|_{q}.\end{equation}As an example this equality does not hold for  scalar products in Hilbert spaces.   For these spaces, \begin{equation}\label{frrqlara}(\frac{r}{q}\langle\psi|\psi\rangle)_{q} \neq \langle\frac{r}{q}\psi|\frac{r}{q}\psi\rangle_{q}.\end{equation}

              \section{Local mathematics and number scaling fields}

              As was noted gauge theories make  use of local vector spaces in space time. Relations between vectors in the different spaces are described by  unitary gauge transformations between vector spaces at different locations.  Here the  existence of local vector spaces is extended to   include structures of numbers of different types.  Locality can be further  extended to mathematical systems of many different types, in particular those whose description includes numbers.

              Here the main emphasis is on vector spaces and real or complex numbers.  For vector spaces and complex numbers, this leads to the existence  of $\bar{C}_{y}\times\bar{V}_{y}$ as complex number structures and vector spaces at each location, $y,$ of a manifold, $M.$  Here $M$ will be taken to be the space time of special relativity although the results are  applicable to more general manifolds.

              In gauge theory the "no information at a distance" principle \cite{Montvay} has the consequence that the information used to choose a basis in a vector space at one location does not determine the choice of a basis in vector space at a different location. Here this principle is extended to  real and complex numbers The result is that the information used to determine a value of a number at one location does not determine the value of the same number at a different location.

              This is implemented here  by first introducing  a real scaling or number value field, $g$.  For each $y$ in $M$, $g(y)$ is the scaling or value factor for the structures at $y.$ This field will be used to introduce a connection as a number preserving value changing map between numbers in structures at different locations.

              Before describing the connection it is worthwhile to describe the collection of scaled complex number structures and vector spaces at each point of space time.  For each point $y$ these  are given by\begin{equation}\label{Fgy}F^{g(y)}_{y}=\bar{C}^{g(y)}_{y}\times\bar{V}^{g(y)}_{y}.\end{equation}  The collection of the $F^{g(y)}_{y}$ for all points, $y,$ in $M$  with the scalars real or complex number structures is the mathematical arena for  much of the rest of this work.

              A good representation of this mathematical arena is by a fiber bundle, $\mathfrak{MF}^{g}$ where \begin{equation}\label{MFg}\mathfrak{MF}^{g}=M\times F, \pi_{g},M\end{equation} The fiber $F$ is defined by \begin{equation}\label{Fbgcp}F=\bigcup_{r}F^{r}\end{equation} where \begin{equation}\label{FrVC}F^{r}=(\bar{C}^{r}\times\bar{V}^{r}).\end{equation} The parameter $r$ varies over all real numbers.  The definition of the projection $\pi_{g}$ differs from that in the usual description of fiber bundles.\cite{FIb}. Here $\pi_{g}$ is a bijective map from $M$ and a subset of the fiber, to $M$.  It has the property such that for each $y$ in $M$ \begin{equation}\label{pim1g} \pi^{-1}_{g}(y)=F^{g(y)}_{y}.\end{equation}  The fiber, $F^{g(y)}_{y},$ is given by Eq. \ref{Fgy}. Also $\pi^{-1}_{g}$ is the inverse of $\pi_{g}.$

              \section{Effects of local mathematics and value fields on physics and geometry}

              The fiber bundle of local complex number and vector space structures with the value field $g$ provides a good setting for the effect of local mathematical structures and the value field on many quantities in physics and geometry. The effect is seen for all quantities and system properties that are represented by integrals or derivatives over space or space time. Examples include fields and wave functions in physics and paths, membranes, and solids in geometry.  Here two generic examples are given, one for integrals and the other for derivatives. Specific examples, gauge theory for physics and paths in geometry, will then be summarized.

              Let $\psi$ be a complex scalar field  on $M.$ For each location $y$ in $M$, $\psi(y)$ is a complex number.  This is the usual representation  of $\psi$ as a function with domain $M$ and range in a single complex number structure. The distinction between $\psi(y)$ as a number and as a number value is not present,

              This description is changed here to describe a field as a section on the fiber bundle, $\mathfrak{MF}^{g}.$  For each $y$ in $M$  $\psi(y)$ is the value of a number in $\bar{C}^{g(y)}_{y}.$  The number with value $\psi(y)$ in $\bar{C}^{g(y)}_{y}$ is denoted by  $\psi(y)_{g(y)}.$

              For the description of integrals it suffices to let $M$ be three dimensional Euclidean space.  The integral of  $\psi(y)_{g(y)}$ over $M$  is not defined because the definition of an integral as the limit of sums of integrands for  different $y$, is not defined.  Arithmetic operations are defined only within local structures, not between structures at different locations.

              This is remedied by the use of a connection to parallel transform a number in $\bar{C}^{g(y)}_{y}$ to the same number in $\bar{C}^{g(x)}_{x}$ at an arbitrary reference location, $x$.  The action of a connection, $C_{g}(x,y)$, on $\psi(y)_{g(y)}$ is given by
              \begin{equation}\label{Cgyxps} C_{g}(x,y)\psi(y)_{g(y)}=(\frac{g(y)}{g(x)})_{g(x)}\psi(y)_{g(x)}=(\frac{g(y)}{g(x)}\psi(y))_{g(x)}.\end{equation} In this equation $(g(y)/g(x))_{g(x)}$ is the number in $\bar{C}^{g(x)}_{x}$ with value $g(y)/g(x),$ and $\psi(y)_{g(x)}$ is a number in $\bar{C}^{g(x)}_{x}$ with value $\psi(y).$ A result of the presence of the $g$ field is that $\psi(y)_{g(x)}$ is a different number than $\psi(y)_{g(y)}.$

              The  desired integral can now be defined as \begin{equation}\label{Ixps}I_{x}(\psi)= (\frac{1}{g(x)})_{g(x)}\int(g(y)\psi(y)dy)_{g(x)}.\end{equation} This integral is defined because the integrands, $(g(y)\psi(y))_{g(x)}$  are all numbers in $\bar{C}^{g(x)}_{x}.$ Also $d(y)$ is an infinitesimal number value of a small number in $\bar{C}^{g(x)}_{x}.$  The factor $g(x)$ is outside the integral as it is independent of $y.$

              The description for scalar fields in the presence of $g$ extends in a straightforward way to vector fields.  If $\psi$ is a field of vector values, then, for each $y$, $\psi(y)_{g(y)}$ is a vector in $\bar{V}^{g(y)}_{y}.$ Parallel transport of these vectors to a reference location, $x$, gives a vector field with all values as vectors in $\bar{V}^{g(x)}_{x}.$ The integral of the resulting localized field is also expressed by Eq. \ref{Ixps}.

              The description of derivatives also shows the effects of the $g$ field. Here $M$ is assumed to be space time of special relativity.  Let $\psi$ be a complex number value field where for each $y$, $\psi(y)_{g(y)}$ is a number in $\bar{C}^{g(y)}_{y}$ with value $\psi(y).$ The partial derivative of $\psi(y)_{g(y)}$ in the direction $\mu=0,1,2,3$ is given by \begin{equation}\label{parpsi}\partial_{\mu,y}\psi(y)_{g(y)}=\lim_{d^{\mu} y\rightarrow 0}\frac{\psi(y+d^{\mu}y)_{g(y+d^{\mu}y)}-\psi(y)_{g(y)}}{d^{\mu}y}\end{equation}.  This expression is not defined. The reason is that $\psi(y+d^{\mu}y)_{g(y+d^{\mu}y)}$ is a number in $\bar{C}^{g(y+d^{\mu}y)}_{y+d^{\mu}y}$ and $\psi(y)_{g(y)}$ is a number in $\bar{C}^{g(y)}_{y}.$ Subtraction is not defined between numbers in different structures.

              As was the case for integrals, this problem is fixed by parallel transport of the number, $\psi(y+d^{\mu}y)_{g(y+d^{\mu}y)}$ to the same number in $\bar{C}^{g(y)}_{y}.$  The resulting partial derivative can be expressed by \begin{equation}\label{parpsiC}\begin{array}{l} D_{\mu,y}\psi(y)_{g(y)}= \mbox{\Large$\frac{C_{g}(y,y+d^{\mu}y)\psi(y+d^{\mu}y)_{g(y+d^{\mu}y)}-\psi(y)_{g(y)}}{d^{\mu}y}$} \\\\\hspace{2cm} = \mbox{\Large$\frac{(\mbox{\large$\frac{g(y+d^{\mu}y)}{g(y)}$}\psi(y+d^{\mu}y))_{g(y)}-\psi(y)_{g(y)}}{d^{\mu}y}$}.
              \end{array}\end{equation} The limit $d^{\mu}y\rightarrow 0$ is implied.

              Taylor expansion of $g(y+d^{\mu}y)$ to first order in differentials gives \begin{equation}\label{gydmuyd}g(y+d^{\mu}y)\simeq g(y) +d_{g}^{\mu}y\partial_{\mu,y}g(y).\end{equation} Use of this in Eq. \ref{parpsiC} gives \begin{equation}\label{pmuygy}
              D_{\mu,y}\psi(y)_{g(y)}=(\partial_{\mu,y}+(\frac{\partial_{\mu,y}g(y)}{g(y)})_{g(y)})\psi(y)_{g(y)}\end{equation} as the final result.

              Eq. \ref{pmuygy} also holds for $\psi$ a vector value field with $\psi(y)_{g(y)}$ a vector in $\bar{V}^{g(y)}_{y}$ with vector value $\psi(y).$  This will be used in the next section where an example from physics and an example from geometry are discussed. The examples will show how local mathematics and the presence of a value field affect theoretical descriptions of quantities in physics and geometry.

              \section{Examples}

              Two examples will be discussed.  One will be the effect of the value field on Dirac Lagrangians in gauge theory.  The other example will be the effect of the value field on distances in geometry.  For both examples the manifold, $M$ will be $4$ dimensional space time as Minkowski space. Local mathematical structures associated with each point of $M$ are assumed.

              It is also useful to replace the value field $g$ by its exponential equivalent, as in \begin{equation}\label{gyeal}g(y)=e^{\alpha(y)}.
              \end{equation} The real scalar field, $\alpha,$ will also be referred to as a scaling or value field.

              In the following the subscripts, $g(y)$ and $g(x),$ will be suppressed.  This makes the mathematical expressions easier to read and more connected to the usual expressions of gauge theory and geometry.  Subscripts will be used where needed to provide clarity.
              \subsection{Gauge theory}

              The Lagrangian density value for Dirac field values is given by \begin{equation}\label{LagD}\mathcal{L}(\psi(y),\partial_{\mu,y}\psi(y)) =\bar{\psi}(y)i\gamma^{\mu}\partial_{\mu,y}\psi(y)-m\bar{\psi}(y)\psi(y).\end{equation} Here $m$ is the mass of the field $\psi$ and
              \begin{equation}\label{bpsi}\bar{\psi}=\gamma^{5}\psi^{*}\end{equation} where $\gamma^{\mu}$ and $\gamma^{5}$ are the gamma matrices \cite{Cheng}.  The corresponding Lagrange density for   Dirac fields is given by\begin{equation}\label{LagDg}(\mathcal{L}(\psi(y),\partial_{\mu,y}\psi(y)))_{g(y)} =(\bar{\psi}(y)i\gamma^{\mu})_{g(y)}\partial_{\mu,y}\psi(y)_{g(y)}-(m\bar{\psi}(y)\psi(y))_{g(y)}.\end{equation}

              Here the description will be given for Abelian gauge theory.  The  fermion  fields, $\psi(y)_{g(y)}$ and $\bar{\psi}(y)_{g(y)}$, are taken to be sections on the fiber bundle, $\mathfrak{MF}^{g}$ with $\psi(y)$ and $\bar{\psi}(y)$  values of the vectors, $\psi(y)_{g(y)}$ and $\bar{\psi}(y)_{g(y)}$, in $\bar{V}^{g(y)}_{y}.$

              So far parallel transport of numbers has been limited to changes in scaling or valuation of scalars  and vectors.  The additional requirement that the Lagrangian density be invariant under local $U(1)$ gauge transformations \cite{Montvay,Cheng}, results in the extension of parallel transport to  apply directly to vectors.

              This is taken care of by expanding the connection, $C_{g}(y,y+d^{\mu}y),$ in Eq. \ref{parpsiC} to include $U(y)=e^{i\phi(y)}$ as a local unitary gauge transformation.  The connection  in Eq. \ref{parpsiC} is replaced by \begin{equation}\label{Calph}
              C_{g}(y,y+d^{\mu}y)=(\frac{e^{\alpha(y+d^{\mu}y)}}{e^{\alpha(y)}}\frac{e^{i\phi(y+d^{\mu}y)}}{e^{i\phi(y)}})_{g(y)}.
              \end{equation}The exponential representation of $g$ is used here. Taylor expansion of both exponents  to first order in $d^{\mu}y$ gives
              \begin{equation}\label{CAiBmu}\begin{array}{l}C_{g}(y,y+d^{\mu}y)\simeq (1+d^{\mu}y(A_{\mu}(y))(1+d^{\mu}yiB_{\mu}(y)))_{g(y)}\\\\\hspace{2cm}\simeq (1+d^{\mu}y(A_{\mu}(y)+ iB_{\mu}(y)))_{g(y)}.\end{array}\end{equation}

              Use of this in Eq. \ref{parpsiC} gives an expansion of Eq. \ref{pmuygy} to include the effects of the $B$ field.  The result is the expansion of Eq. \ref{pmuygy} to \begin{equation}\label{pmuygyD}D_{\mu,y}\psi(y)=(\partial_{\mu,y}+A_{\mu}(y) +iB_{\mu}(y)))\psi(y),\end{equation}(The subscript, $g(y)$ is suppressed).

              Use of this in the Dirac Lagrangian density value gives \begin{equation}\label{LagDD}\begin{array}{l}\mathcal{L}(\psi(y),\partial_{\mu,y}\psi(y)) =\bar{\psi}(y)i\gamma^{\mu}D_{\mu,y}\psi(y)-m\bar{\psi}(y)\psi(y)\\\\\hspace{1cm}=\bar{\psi}(y)i\gamma^{\mu}(\partial_{\mu,y} +aA_{\mu}(y)+ibB_{\mu}(y))\psi(y) -m\bar{\psi}(y)\psi(y).\end{array}\end{equation}Coupling constants for the $\vec{A}$ and $i\vec{B}$ fields are denoted by $a$ and $b$.  Since $\vec{B}$ is the electromagnetic or photon field, $b$ is the square root of the fine structure constant. The value of $a$ is not known so far.

              The requirement that the Lagrangian density value be invariant under local $U(1)$ transformations is expressed by
              \begin{equation}\label{DpUUD} D_{\mu,y}'U\psi(y)=UD_{\mu,y}\psi(y)\end{equation}If $U=e^{i\theta(y)}$ this equation gives the condition
              \begin{equation}\label{ApA}A'_{\mu}(y)=A_{\mu}(y)\end{equation}and \begin{equation}\label{BpB}B'_{\mu}(y)=B_{\mu}(y)-\frac{1} {b}\partial_{\mu,y} \theta(y).\end{equation}From these equations one concludes  the well known fact that the photon mass is $0.$  The $\vec{A}$ field can have any mass, including $0$.

              The $\vec{B}$ field is nonintegrable.  This is a consequence of the Aharonov Bohm effect \cite{AB}. As defined here the $\vec{A}$ field is integrable. The definition of the connection can be changed  to allow for $\vec{A}$ to be nonintegrable.

              One concludes from this that the $\vec{A}$ field is a real scalar field with any mass possible.  It is also assumed to be a spin $0$ field.
              At present it is not known which physical field, if any, corresponds to the $\vec{A}$ field. The Higgs field is the only spin $0$ scalar field in the standard model. Are the two fields related?

              The $\vec{A}$ field also appears as a real, scalar  field in nonAbelian gauge theories. The presence of the $\vec{A}$ field in these Lagrangians may be of help in deciding which physical field, if any, is represented by $\vec{A}.$

              \subsection{Local scaled geometry}

              The presence of the field $\alpha$ and local mathematics affects many geometric properties.   For these properties the fiber contents are different from those for gauge theories.  Here the fiber bundle has the same form given by $\mathfrak{MF}^{g}.$ The contents of the fiber are different in that\begin{equation}\label{FrT} F=\bigcup_{r}(\bar{R}^{r}\times\bar{T}^{r}).\end{equation} The contents of the fiber at each location of $M$  are given by\begin{equation}\label{piyRT}\pi^{-1}_{g}(y)=F^{g(y)}_{y}=\bar{R}^{g(y)}_{y} \times\bar{T}^{g(y)}_{y}.\end{equation} Here $\bar{R}^{g(y)}_{y}$ is a scaled real number structure and $\bar{T}^{g(y)}_{y}$ is a local scaled representation of $M$ at $y.$

              A good representation for $\bar{T}^{g(y)}_{y}$ is by the quadruple, $\bar{R}^{4,g(y)}_{y}$ of scaled real numbers. The associated metric tensor is $(h_{\mu,\nu})_{g(y)}=((1,-1,-1,-1)\delta_{\mu,\nu})_{g(y)}.$ The space, $\bar{T}^{g(y)}_{y}$ can be defined as a scaled chart representation of $M$.  For each $y$\begin{equation}\label{Trhoygy}\bar{T}^{g(y)}_{y}=\rho^{g(y)}_{y}(M).\end{equation}  Since $M$ is flat, the chart, $\rho^{g(y)}_{y},$ is defined on all of $M$.

               The effect of scaling of local real lumber structures and representations of $M$ can be seen in the example of the length of a path.  Let $p$ be a path on $M$ with beginning and end points, $y$ and $z.$ Let the path be parameterized by a real number $s$ with $p(0)=y$ and $p(1)=z.$  The length value of the path is defined by \begin{equation}\label{Lp}L(p)=\int_{0}^{1}[\partial_{\mu,s}p h_{\mu,\nu}\partial_{\nu,s}p]^{1/2}ds.\end{equation}Sum over repeated indices is implied.

               Under the assumption of global mathematics, this  path integral is well defined. With the assumption of local mathematics and the presence of the value field, there is a vector space structure, $\bar{T}^{g(p(s)}_{p(s)}$ associated with each path point, $p(s),$ on $M$. The path gradients, $\nabla_{s}p,$ form a field of vectors along the path.  Treatment of this field as a section on the fiber bundle, defined by Eqs. \ref{FrT} and \ref{piyRT}, has the consequence that for each $s$,  the vectors with components, $\partial_{\mu,s}p$ are vectors in $\bar{T}^{g(p(s)}_{p(s)}$.

               As a result the  integral  of Eq. \ref{Lp} is not defined.  The reason is that for each $s,$ the integrand is a number in $\bar{R}^{(p(s))}_{p(s)},$ and $\partial_{\mu,s}p$ is the $\mu$ component of a vector in $\bar{T}^{g(p(s))}_{p(s)}.$ The integrand addition implied in the definition of an integral, is not defined between numbers in different structures.  it is defined only for numbers within a structure.

               This problem is fixed by use of a connection to parallel transport the integrands to a real number and position space structure at a reference location $x$.   The transport gives \begin{equation}\label{Cgxps}\begin{array}{l}[\partial_{\mu,s}p h_{\mu,\nu}\partial_{\nu,s}p]^{1/2}_{g(p(s))} \rightarrow C_{g}(x,p(s))[\partial_{\mu,s}p h_{\mu,\nu}\partial_{\nu,s}p]^{1/2}_{g(p(s))} \\\\\hspace{1cm}=(e^{\alpha(p(s))-\alpha(x)})_{g(x)}[\partial_{\mu,s}p h_{\mu,\nu}\partial_{\nu,s}p]^{1/2}_{g(x)}.\end{array}\end{equation}  Subscripts have been added to to show that the factors with subscripts are numbers in $\bar{R}^{g(x)}_{x}$ with the indicated number values.

               The integral of the transported path lengths is defined.  It is given by \begin{equation}\label{Lpgx} L(p)_{x}=(e^{-\alpha(x)} \int_{0}^{1}e^{\alpha(p(s))}[\partial_{\mu,s}p h_{\mu,\nu}\partial_{\nu,s}p]^{1/2}ds)_{g(x)}.\end{equation}  This is the length of the path $p$ from $y$ to $z$ in the presence of the scaling or value field, $\alpha.$ The path length is a real number quantity in $\bar{R}^{g(x)}_{x}.$

               The distance or geodesic between $y$ and $z$ is the length of the shortest path between $y$ and $z$.  This is obtained by the use of the Euler Lagrange equations to find the minimum of $L(p)_{x}$ by variation of the path, $p.$ The result is given by
               \begin{equation}\label{Gdp}[\frac{d}{d\tau}+\vec{A}(p(\tau))\cdot\nabla_{\tau}p]\frac{dp^{\mu}}{d\tau}-h^{\mu,\nu}A_{\nu}(p(\tau))=0.
               \end{equation}  This is the geodesic equation for the component $p^{\mu}$ of the minimal path. The term, $\vec{A}(p(\tau))\cdot\nabla_{\tau}p$  is obtained as $$\vec{A}(p(\tau))\cdot\nabla_{\tau}p=e^{-\alpha(p(\tau))}\frac{d} {d\tau}e^{\alpha(p(\tau))}.$$  Details of the derivation are given in \cite{BenFB}. The proper time is denoted by $\tau.$

               This equation shows that, if all components of $\vec{A}$ are $0$, then Eq. \ref{Gdp}  reduces to \begin{equation}\label{d2tau}\frac{d^{2}p^{\mu}} {d\tau^{2}}=0. \end{equation}This is the equation for a straight line in space time. From Eq. \ref{Gdp} one sees that the presence of the $\vec{A}$ field affects the distance between locations in $M$.  It would be interesting to see if this effect shows up in physics in some way. Dark energy?

               \section{Restrictions on the $\alpha$ field}

               So far there is no direct experimental evidence for the presence of the $\alpha$ field or its gradient, $\vec{A}.$  The great accuracy of quantum electrodynamics shows that either the coupling of the $\vec{A}$ field to fermion fields must be extremely small compared to the fine structure constant, or $\vec{A}$ must be very close to zero. Other areas of physics also show no experimental evidence for the presence of $\vec{A}.$

               This fact is to be combined with an important property of all experiments.  This is the fact that all experiments carried out to date are implemented locally. They are done by us on or close to the surface of the earth.  This includes experiments to determine properties of systems that exist locally as well as properties of cosmological systems such as galaxies and pulsars.

               In the future humans may  inhabit other solar system planets and interplanetary space ships.  It is assumed here that any experiment done by humans in these locations will find no direct experimental evidence for the presence of the $\vec{A}$ field.

               This lack of  experimental evidence for the presence of $\vec{A}$  should also be extended to potential inhabitants of planets around stars that are within effective communication with us.  These observers can describe experiments they have carried out.  They can also communicate to us the results and  their interpretations of these results.  It is assumed here that they also will find no direct evidence for the presence of $\vec{A}.$

               These assumption regarding the lack of direct experimental evidence for the lack of $\vec{A}$ can be combined into a restriction that
               \begin{equation}\label{vcA0}|\vec{A}(y)|<\epsilon \simeq 0\end{equation}  for all space time points, $y,$ in a local region of the universe. Here $|\vec{A}(y)|$ is the length of $\vec{A}(y).$ Within this local region $\epsilon$ is smaller than the uncertainty of all experiments carried out to date and presumably in  the future.

               One may expect that in the future experiments will be done with increasing accuracy in the results. However uncertainties will never be exactly $0.$  At present effective limits are presented by the Planck units of length, time, and mass.  However the possibility that even these limits will be surpassed in the future cannot be ruled out.

               The important aspect of these restrictions is that they are limited to apply only to a local region of the universe.  The size of the local region is determined by the maximum distance between us on the earth, and observers on other planets with whom we can effectively communicate.

               The size of this local region is unknown.  One literature estimate \cite{AmSci} is that it is restricted to those stars that are at most, $1000$ light years distant from us.  Here the size of the restricted region is not important.  The only requirement is that it is a small fraction of the $14$ billion light year size of the universe.

               Outside the local region there are no restrictions on the variation of $\vec{A}.$ It can vary rapidly with location or slowly or not at all.

               There are fields described in physics that share the restrictions of $\vec{A}.$   These include  the inflaton \cite{Infl}, quintessence \cite{Quint}, and dark energy \cite{DE}.\footnote{See \cite{BenFB} for a discussion of the possible relation of $A$ and $\alpha$ to the Higgs field.}  The effects of these fields are present globally.  They are invisible locally. At present it is an open question if $\vec{A}$, or $\alpha,$ are any of these fields, or none of these.  This is a problem for future work.

               \section{Discussion}

                In this paper it was seen that the no information at a distance principle and the assumption of local number and vector structures led to the existence of a number scaling or value field, $g.$ This field had two  roles,  It determined the scaling factor for number and vector structures at each point of space time. It also was used to define the connection between structures at different locations.  On number structures the connection was a number preserving  number value changing operation.  On vector spaces the connection was a vector preserving vector value changing operation.  Its effect was to multiply a vector by a $g$ dependent scalar.

                Theoretical quantities in physics that are defined by integrals or derivatives of some field or function are affected by the presence of $g$. This was seen from the need to parallel transport values of scalar or vector fields in different structures at different locations  to a common location.  Integrals and derivatives of sections of fields so transported were seen to be well  defined.    As a result theoretical descriptions of many quantities in physics are affected by the presence of the $g$ field, or its equivalent, the $\alpha$ field where $g(y)=e^{\alpha(y)}.$

                Two examples of  effect of $\alpha$ on physical quantities were given.   One example showed the effect on the Dirac Lagrangian in gauge theory.  The other example showed the effect on path lengths and distances. It was seen that the presence of the $\alpha$ field resulted in an   additional vector field in the Lagrangian that interacts with the fermion field. The same vector field appears in the geodesic equation for the distance between two  locations in space time. The vector field is the gradient of the scalar $\alpha$ field.

                Other examples of the effect of the $g$ or $\alpha$ field on physics and geometry are given in the cited references. These include the description of black and white holes in geometry where $\alpha(y)$ approaches either $+$ or $-$ infinity as $y$ approaches a point, $x$ \cite{BenNOVA}  and the effect of $\alpha$ on quantum mechanical quantities \cite{Ben2nd}.

                There is still much to do. The effect of number scaling or valuation needs to be extended to mathematical structures  other than scalars and vector spaces.  Examples are the operator algebras much used in quantum mechanics and the Lie algebras and matrices for unitary transformations used in gauge theories.  In geometry the description for paths needs to be extended to two,  three, and four  dimensional systems.  The effect also needs to be expanded to include general relativity. The relation of each fiber with greatly expanded mathematical content to the mathematics that is potentially available to a conscious observer at the location of a fiber is intriguing.  This should also be investigated.

               \section*{Acknowledgement}
                This material is based upon work supported by the U.S. Department of Energy, Office of Science, Office of Nuclear Physics, under contract number DE-AC02-06CH11357.


\begin{thebibliography}{99}

                \bibitem{Wigner}
            E.  Wigner,  "The unreasonable effectiveness of mathematics in the natural sciences", Commun.  Pure Appl. Math. \textbf{13}, No. 1, (1960).  Reprinted in E. Wigner,
           \emph{Symmetries and Reflections}, (Indiana Univ. Press, Bloomington, IN 1966), pp222-237.

             \bibitem{Omnes}
           R. Omnes,   "Wigner's Unreasonable Effectiveness of Mathematics", Revisited,
           Foundations of Physics, \textbf{41}, 1729-1739, (2011).

            \bibitem{Plotnitsky}
            A. Plotnitsky, On the reasonable and unreasonable effectiveness of mathematics in
            classical and quantum physics, Foundations of Physics, \textbf{41}, pp. 466-491, (2011).

            \bibitem{Hitchin}
            N. Hitchin, "Interaction between mathematics and physics", Arbor Ciencia, Pensamento y cultura, CLXXXIII \textbf{725}, pp. 427-432, (2007).


            \bibitem{Dirac}
            P. A. Dirac, "The relation between mathematics and physics", Proc. Royal Soc. of Edinburgh, \textbf{59}, Part II, pp. 122-129, (1938).


            \bibitem{Von Neumann}
            J. von Neumann, "The works of the mind",\textbf{1}, 1:pp 180-196, (1947).



            \bibitem{WikiRMP}
            Wikipedia: "Relationshhip between Mathematics  and physics" See for these and additional references.

            \bibitem{BenCTPM1}
            P. Benioff, Towards a Coherent Theory of Physics and Mathematics
            Foundations of Physics, \textbf{32},  pp. 989-1029, (2002):   arXiv:quant-ph/0201093


            \bibitem{BenCTPM2}
            P. Benioff, Towards a coherent theory of physics and mathematics: the theory-experiment connection,  Foundations of Phys., \textbf{35}, 1825-1856, (2005).

            \bibitem{Shapiro}
             Shapiro, S.: Mathematical Objects, in Proof and other dilemmas, Mathematics and philosophy, Gold, B. and  Simons, R., (eds), Spectrum Series, Mathematical Association of America, Washington DC, 2008, Chapter III, pp 157-178.

            \bibitem{Montvay}
            I. Montvay and G. M\"{u}nster, Quantum fields on a lattice, Cambridge University Press, UK,(1994), Chapter 3.

             \bibitem{Mack}
             G. Mack,  "Physical principles, geometrical aspects, and locality properties of gauge field theories," Fortshritte der Physik, \textbf{29}, 135, (1981).

                \bibitem{YangM}
            C. N. Yang and  R. L. Mills,  Conservation of Isotopic Spin and Isotopic Gauge Invariance, Phys. Rev., \textbf{96}, 191-195,  (1954)

              \bibitem{BenNOVA}
             P. Benioff,  "Gauge theory extension to include number scaling by boson field: Effects on some aspects of physics and geometry," in \emph{Recent Developments in Bosons Research}, I. Tremblay, Ed., Nova publishing Co., (2013), Chapter 3;  arXiv:1211.3381.


            \bibitem{Ben1st}
            P. Benioff, New Gauge Field from Extension of Space Time Parallel Transport of Vector Spaces to the Underlying Number Systems
             International Journal of Theoretical Physiocs,\textbf{ 50}, pp. 1887-1907 (2011): arXiv:1008.3134

             \bibitem{BurginDio}
             M. Burgin, Diophantine and nondiophantine arithmetics: operations with numbers in science and everyday life. arXiv:math/0108149.

             \bibitem{BurginDio2}
             M. Burgin, Nondiophantine arithmnetics or is it possible that 2+2 is not equal to 4.
             Ukrainian Academy of information sciences, Kiev,1997 (in Russian).

             \bibitem{Burginpro}
             M. Burgin, Introduction to projective arithmetics, arxiv:1010.3287.

                 \bibitem{Czachor1}
             M. Czachor, "Relativity of arithmetics as a fundamental symmetry of physics", Quantum Studies: Mathematics and Foundations, \textbf{3}, 123-133, (2016): arXiv:1412.8583

              \bibitem{Czachor2}
               M. Czachor, Dark energy as a manifestation of nontrivial arithmetic
                Int. J. Theor. Phys. \textbf{56}, pp. 1364-1381 (2017) :arXiv:1604.05738


              \bibitem{BenINTECH}
             P. Benioff, "Effects on quantum physics of the local availability of mathematics and space time dependent scaling factors for number systems", in \emph{Quantum Theory}, I. Cotaescu, Ed.,  Intech open access publisher, 2012, Chapter 2, arXiv:1110.1388.

              \bibitem{BenFB}
               P. Benioff, Fiber bundle description of number scaling in gauge theory and geometry
               Quantum Studies: Mathematics and Foundations, 2, 289-313,( 2015), arXiv:1412.1493

             \bibitem{Ben2nd}
              P. Benioff,  Effects of a scalar scaling field on quantum mechanics,
              Quantum Information Processing, \textbf{15}(7), pp. 3005-3034, (2016), arXiv:1512.05669


                \bibitem{BurginNamed}
                M. Burgin, Unified foundations for mathematics, arxiv:math/0403186.


                 \bibitem{FIb}
                D. Husem\"{o}ller, \emph{Fibre Bundles}, Second edition,  Graduate texts in Mathematics, v. 20, Springer Verlag, New York, (1975).

                \bibitem{Cheng}
                T. P. Cheng and  L. F. Li, \emph{Gauge Theory of Elementary  Particle Physics}, Oxford University Press, Oxford, UK, (1984), Chapter 8.

                \bibitem{AB}
                Y. Aharonov and D. Bohm, Phys. Rev. \textbf{115}, 485, (1959).


               \bibitem{AmSci}
           H. A. Smith, "Alone in the Universe", American Scientist, \textbf{99},  No. 4, p. 320, (2011).

             \bibitem{Infl}
             F. Bezrukov and M. Shaposhnikov,  The standard model Higgs boson as the inflaton,  Physics Letters B, \textbf{659},703-706, (2008).

              \bibitem{Quint}
              I. Zlatev, L. Wang, L. P. Steinhardt,   Quintessence, Cosmic Coincidence, and the Cosmological Constant, Physical Review Letters \textbf{82} (5): 896–899, 1999; arXiv:astro-ph/9807002.


             \bibitem{DE}
             Miao Li, Xiao-Dong Li, Shuang Wang, Yi Wang,
             Dark energy, a brief review,  Frontiers of Physics, \textbf{8},  pp. 828-846, (2013)










              \end{thebibliography}
           \end{document}